# Seedless assembly of colloidal crystals by inverted micro-fluidic pumping

Ran Niu[*], Thomas Palberg

*Institut für Physik, Johannes-Gutenberg Universität, Staudingerweg 7, 55128, Mainz*

**Abstract**

We propose a simple seedless approach to assemble millimeter sized monolayer single colloidal crystals with desired orientations at predetermined locations on an unstructured charged substrate. This approach utilizes the millimeter-ranged fluid flow on the bottom glass substrate induced by an ion exchange resin (IEX) fixed on top of the closed sample cell. This fluid flow increases with decreasing height of the sample cell and increasing radius $R$ of the IEX. For a single inverted pump, millimeter sized monolayer single crystals of hexagonal close packing can be obtained. For two closely spaced ($D \sim 4R$) pumps, the formed crystals have a predefined orientation along the line connecting the two IEX. By patterning IEX into different structures, colloidal crystals of different complex patterns form. The present method paves a convenient way for fabricating high quality monolayer colloidal crystals for a variety of applications.

**Introduction**

Colloidal suspensions of spherical particles self-assemble into crystalline structures once their interaction and density are in the suitable range.[1-4] These crystalline structures have achieved great attention motivated by their unique optical properties and potential applications as photonic or phononic materials[5,6] as well as their usage as model system for addressing fundamental problems in soft condensed matter.[2,7] In the past two decades, great effort has been devoted to fabricating large sized single crystals (2D and 3D) of defined orientations or at pre-determined locations.[8] The fabrication approaches rely on evaporation,[9,10,11] electric fields,[12] electro-convection,[13] gradients, e.g., in temperature[14-16] or electrolyte,[17] or shear flow.[18-21] However, realizing simultaneous position and orientation control is still challenging. Successful approaches so far rely on template assisted growth, e.g., topological templates[22,23] or light grids.[24,25]

With the rapid development of active matter, a new approach relying on the flow generated by catalytic, thermo-phoretic, diffusio-osmotic, or electro-osmotic (eo) pumps to assemble colloidal particles has been proposed.[26-30] This approach has shown its advantage in transporting, assembling and delivering cargo on demand.[31-33] However, most of these pumps work at elevated electrolyte concentrations or even at physiological conditions, making it hard to assemble charged colloids at dilute concentrations. Moreover, the formed crystals are often of low quality, showing many defects (multidomains and dislocations).[26,32] The first experimental realization of colloidal assembly at low ionic level was observed beneath an ion exchange resin (IEX) about 40 years ago.[34] However, the mechanism for the assembly was not clear at that time. In our previous work, we demonstrated that the ion exchange at micromolar salt concentration leads to diffusio-electric fields, which act on a charged substrate inducing a converging eo-flow. The eo-flow drags colloids to the vicinity of the IEX.[35] To obtain high quality crystals, we further coated the negatively charged substrate with positively charged surfactants, which decreases the surface charge and thus the flow strength.[36] The competition between the counterbalancing electrophoretic and electro-osmotic flows leads to the sorting of colloidal particles according to their size. Furthermore, it was proved that a finite assembly distance between the colloidal crystal and the IEX is crucial for the formation of high quality crystals. However, for crystals formed around a seed, one common problem is the incommensurability of the seed and the formed crystal, which usually leads to defects in the crystal (Fig. S1 in the supporting information).

In this work, we modify our fabrication approach by simply placing the IEX on top of the sample cell and utilizing the fluid flow on the bottom substrate to assemble colloidal crystals. This simple modification benefits our approach from the following points. First, colloidal particles directly assemble underneath the IEX hanging above the bottom substrate, which does not have seeds, thus the formed crystals have no voids in the center. Second, this allows easy control of the range and strength of the flow by tuning the cell height and the size of the IEX without surface coating, as long-ranged weak flow facilitates the formation of large sized high quality crystals. Third, this assures easy control of the flow pattern by changing the configurations of the IEX on the top substrate, therefore the colloidal crystals can be deliberately oriented and shaped.

In the following, we first present the quantification of the fluid flow generated by differently sized IEX and at different cell heights, followed by analyzing the monolayer colloidal crystals formed in single IEX- and two IEX-based inverted pumps. Then we show the first attempts of patterning colloidal crystals in inverted pumps. Finally, we discuss the application range, limitations and future applications of the inverted pump-guided colloidal assembly.

**Results**

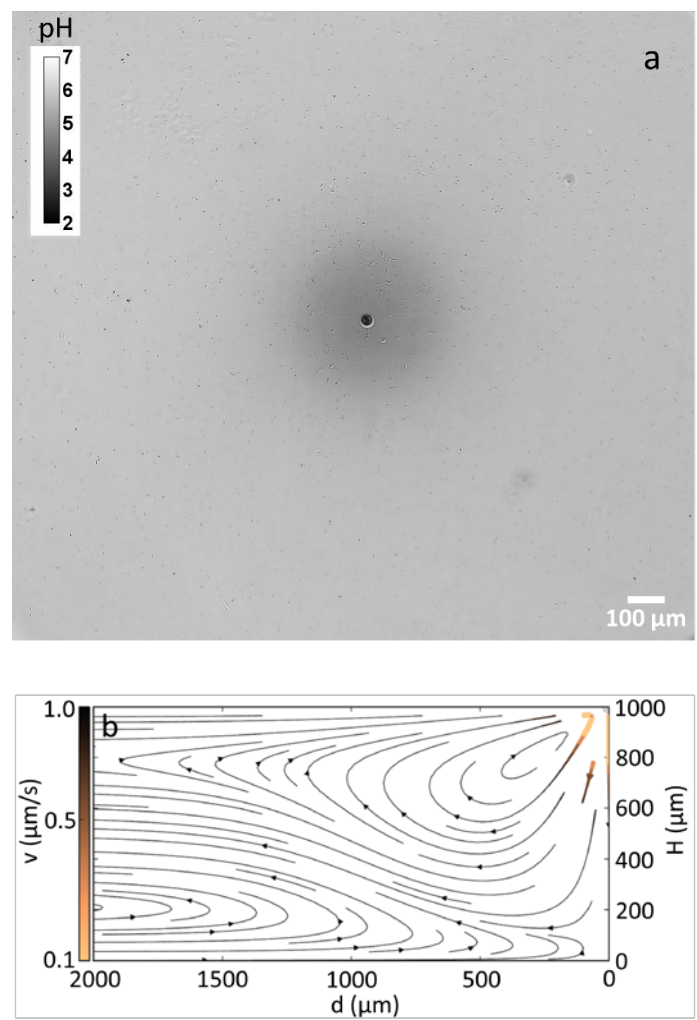

**Fig. 1** a) pH gradient image of an IEX45-based inverted pump with pH values indicated by the calibration bar. b) Visualization of the FEM calculated fluid flow generated by an IEX45-based inverted pump at cell height $H$= 1 mm over a radial range of 2 mm (full range of simulation = 3 mm).[35] The IEX45 is fixed in the right-top corner, the direction of flow is shown with arrows, and the color indicates the magnitude of the local velocity.

The mechanism of the IEX based eo-pump has been demonstrated in our previous work.[35] Generally, the IEX exchanges micromolar concentration of impurity cations for stored protons and generates a concentration gradient that decays from the particle. Using the photometry method developed in our group,[37] we have quantified the local pH gradient (Fig. 1a). Different diffusion coefficients of impurity cations and protons induce diffusio-electric ($E$) fields pointing inward. The $E$ fields act on the negatively charged substrate inducing a converging eo-flow, which is 2-dimensional in a confined cell and 3-dimensional for an unconfined cell. Finite element method (FEM) simulation solves the 3-dimensional hydrodynamics revealing a mirrored converging flow on the top substrate and a back flow at intermediate heights due to the closed cell geometry.[35] Since both the ion exchange and the flow are not influenced by gravity, an inverted geometry can be constructed, where the IEX is fixed on top of the sample cell. For the IEX45-based inverted microfluidic pumps, the fluid flow is shown in Fig. 1b. For clarity, a schematic drawing of the IEX-based inverted microfluidic platform is shown in Fig. S2.

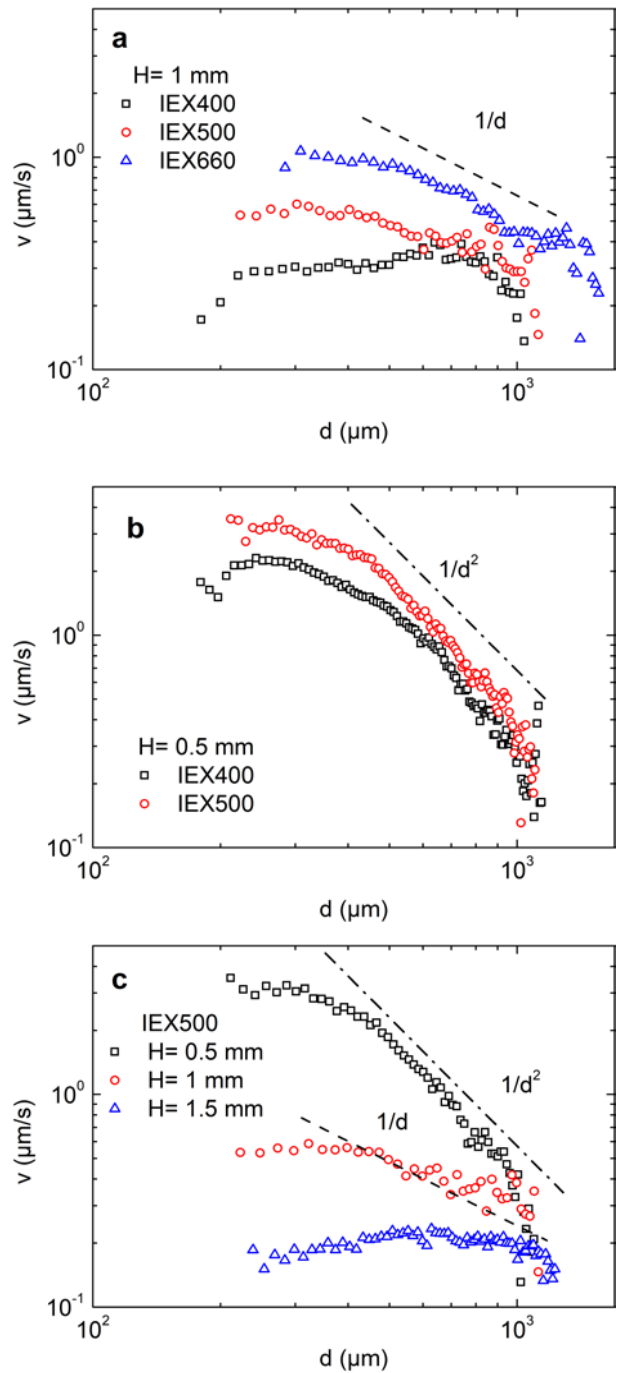

**Fig. 2** Fluid flow on the bottom substrate as a function of distance *d* to the center of IEX for differently sized IEX at cell height *H*= 1 mm (a), *H*= 0.5 mm (b) and for IEX500 at different cell heights (c).

To get a significantly long-ranged fluid flow on the substrate, we used large sized IEX compared to that used in the normal eo pumps. Using tracer velocimetry method, we detect the fluid flow on the bottom substrate for single IEX-based inverted pumps (video 1). The flow at different cell heights and for differently sized IEX is shown in Fig. 2. For these inverted pumps, the fluid flow is in the radial range of about 1 mm, which facilitates colloidal assembly from a large area. At the same cell height, the strength of the fluid flow increases as the size of IEX increases (Figs. 2a and 2b). At $H$= 1 mm, the fluid flow is in the range of 0.3~1 µm/s. Moreover, in the radial range of 400-1000 µm the flow still obeys the $1/d$ dependence, typical of 2-dimensional flow. However, when the size of the IEX is smaller than half the cell height, the $1/d$ dependent region dramatically shrinks. Interestingly, when the size of the IEX is close to the cell height, the fluid flow shows $1/d^2$ dependence in the radial range of 400 µm $< d <$ 1000 µm, typical of 3-dimensional flow. We will shortly discuss this point below. For the IEX500-based inverted pumps, the fluid flow decreases to ~1/3 and ~1/10 when the cell height is increased from $H$= 0.5 mm to $H$= 1 mm and $H$= 1.5 mm, respectively (Fig. 2c). To obtain a weak flow for high quality crystal formation within experimental time, most of the results shown below were obtained from IEX500-based inverted pumps at $H$= 1 mm, except otherwise noted.

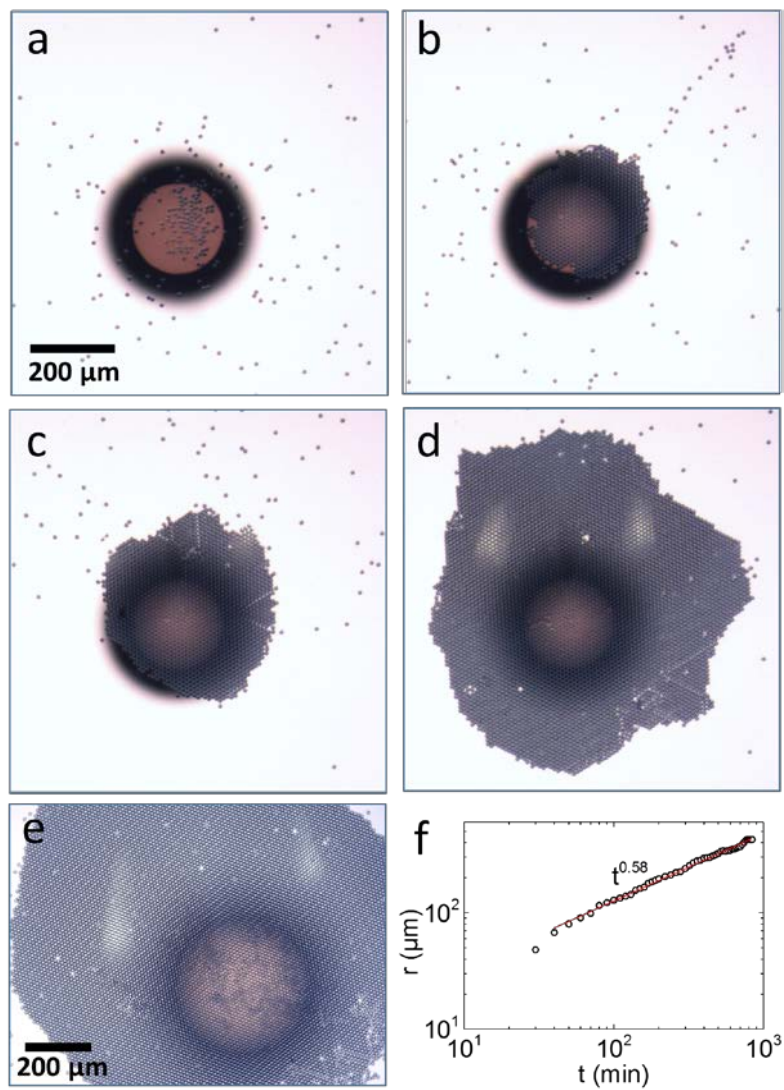

**Fig. 3** Typical stages of single colloidal crystal formation: PS15 in the single IEX500-based inverted pump (a-e). Note the different scale bar used for (e). Crystal radius *r* as a function of time from the start of the experiment (f).

Fig. 3 shows the typical stages of single colloidal crystal formation in single IEX-based inverted pumps (Fig. 3 and video 2). PS15 particles are dragged by the fluid flow towards the projected center of the IEX500. As more particles get accumulated, these particles form hexagonal close packing structure beneath the IEX (Fig. 3a) and the crystal grows larger (Figs. 3b-e). In the crystal growth process, grain boundaries or dislocations anneal spontaneously (video 2). The crystal radius *r* increases with $t^{0.58}$, indicating a transport controlled colloidal assembly (Fig. 3f). Within 11 h, the monolayer single crystal can grow up to millimeter size (Fig. 3e). Owing to incomplete annealing or slight size polydispersity of particles, there are a few defects (vacancies, grain boundaries and dislocations) left in the crystal. The defect fraction is ~1.0%. Measuring the crystal growth at several arbitrarily chosen locations of the IEX in the sample cell, no difference of the formed crystals was detected. In fact, the mechanism of assembly works independently of the pump location as long as a sufficient distance to the cell boundary is maintained. Hence, we can grow crystals at arbitrarily predetermined positions.

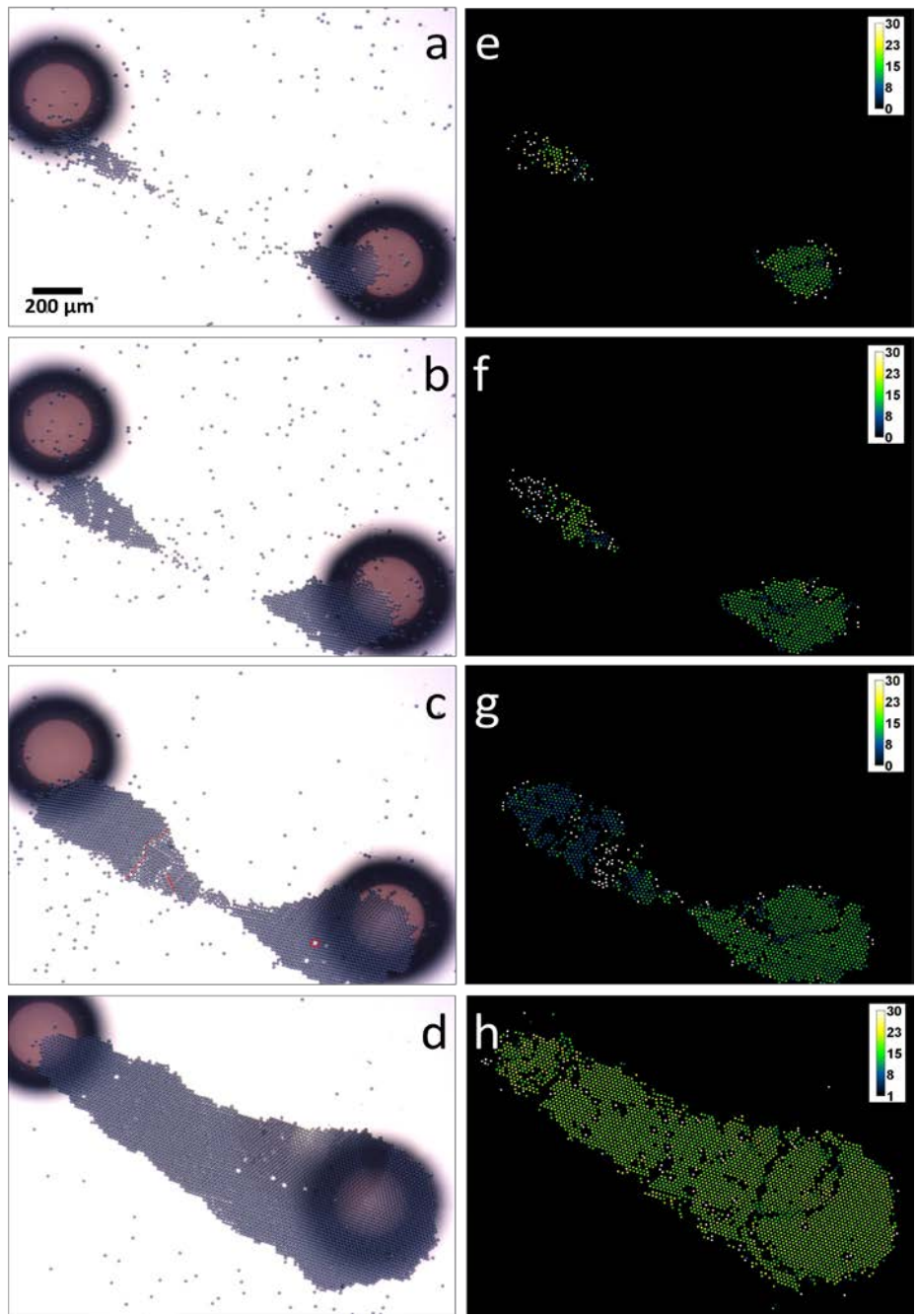

**Fig. 4** Typical stages of oriented crystal formation: PS15 in the inverted pump formed by two spaced IEX500 (center-to-center distance of 1487 μm, aligned along 30° relative to the horizontal axis). 200 μm scale bar shown in (a) applies for all images. (a) At $t$ = 42min, crystals form beneath the two IEX500. (b) At $t$ = 72min, crystals grow parallel to the line connecting the two IEX500. (c) At $t$ = 102 min, the two crystals get in touch and anneal. Examples of three kinds of defects: vacancy (red circle), dislocation (red solid line) and grain boundary (red dashed line) are labelled. (d) At $t$ = 166 min, a large monolayer single crystal of single orientation forms along the line connecting the two IEX500. The corresponding

orientation maps of the crystalline particles with $\phi_6 > 0.8$ and the color coded local crystal orientation in ° relative to the horizontal axis (e-h).

It is also possible to manipulate the orientation of colloidal crystals. In Fig. 4 and video 3, we place two IEX500 on top of the sample cell, which align along 30° relative to the horizontal axis and with center-to-center distance of 1487 μm. As shown in Fig. 4a, two crystals form beneath the two IEX. As the PS15 particles are dragged to the space between the two IEX500, the crystals grow parallel to the central line (Fig. 4b). After the two crystals get in touch, they anneal (Fig. 4c). Then the crystal grows in the direction perpendicular to the central line (Fig. 4d). In the whole process, the dislocations or grain boundaries develop and anneal (video 3). The orientation of crystalline particles relative to the horizontal axis is quantified for bond order parameter $\phi_6 > 0.8$ (Figs. 4e-h). Note that, the existence of IEX induced inhomogeneous illumination results in slight imperfections in particle detection (several particles cannot be detected), especially around the edge of the IEX.

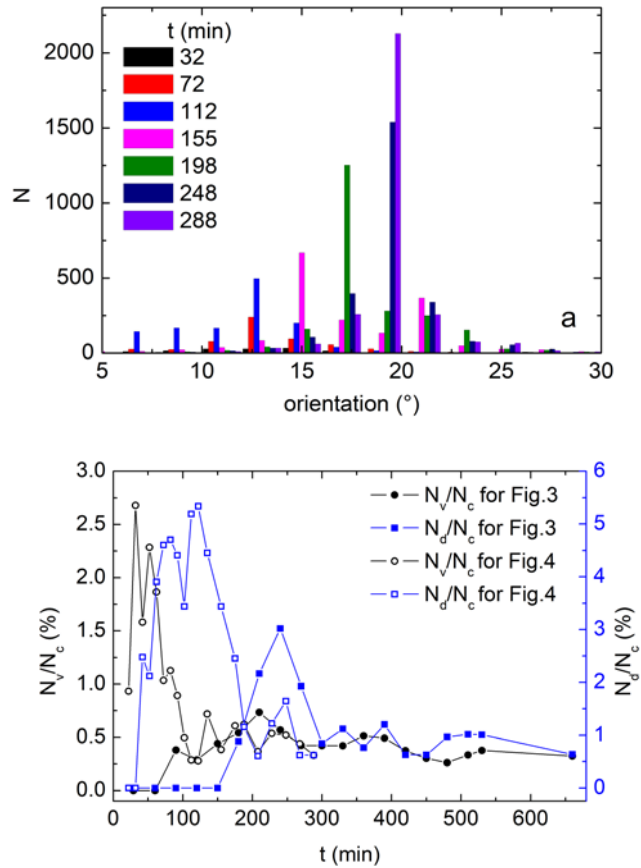

**Fig. 5** (a) Histogram of the number of crystalline particles with certain orientations relative to the horizontal axis. (b) Fraction of voids and vacancies $N_v/N_c$ (left scale, open circles) and fraction of particles in dislocations and grain boundaries $N_d/N_c$ (right scale, open squares) as a function of time $t$ from the start of the experiment. For comparison, we also show the values obtained for the crystal shown in Fig. 3, with $N_v/N_c$ labeled with solid circle and $N_d/N_c$ labeled by solid squares.

The histogram of number of crystalline particles with certain orientations is shown in Fig. 5a. The orientation of individual grains of crystals formed in the beginning is fairly random. After annealing, the distribution of orientations gets narrower and the orientation peak shifts from 12.5° to 20° relative to the horizontal axis, which is approaching the orientation of the line connecting the two IEX500 (30°). When the distance between the IEX500 is decreased to 1116 μm or even smaller, single colloidal crystal forms and grows in the space between the IEX500 (Fig. S3 and video 4). Nevertheless, the final orientation of the single crystal is close to that of the line connecting the two IEX500 (Figs. S3 and S4, video 4). The fraction of voids and vacancies $N_v/N_c$ and the fraction of particles in dislocations and grain boundaries $N_d/N_c$ can be quantified (Fig. 5b and Fig. S4b). Initially, $N_v/N_c$ rapidly increases as two loose crystals form. It then steadily decreases as the crystal size increases with some fluctuations resulting from the development and annealing of dislocations and grain boundaries. Dislocations and grain boundaries occur at later stages. $N_d/N_c$ shows a pronounced peak during the main annealing stage. Some slower and weaker fluctuations seen in $N_d/N_c$ seem to be correlated to the flow induced orientation change of the crystal. Finally, a single crystal with only less than 1% of defects and defined orientation close to that of the line connecting the two IEX has formed. For comparison, we also plot the change of $N_v/N_c$ and $N_d/N_c$ with time for the crystal formed in single IEX-based inverted pump (Fig. 3). The two parameters have similar developing trend with that in two IEX-based inverted pumps. However, the peaks of $N_v/N_c$ and $N_d/N_c$ are lower and appear at later time for crystal formed in single IEX-based pump than that in two IEX-based pumps, indicating faster annealing of crystal in two IEX-based inverted pumps. We note that the defect fraction can be further decreased by more careful preparation of samples, e.g., separating impurities (ion exchange debris etc.) from the colloidal suspension and plasma cleaning the substrate after chemical cleaning.

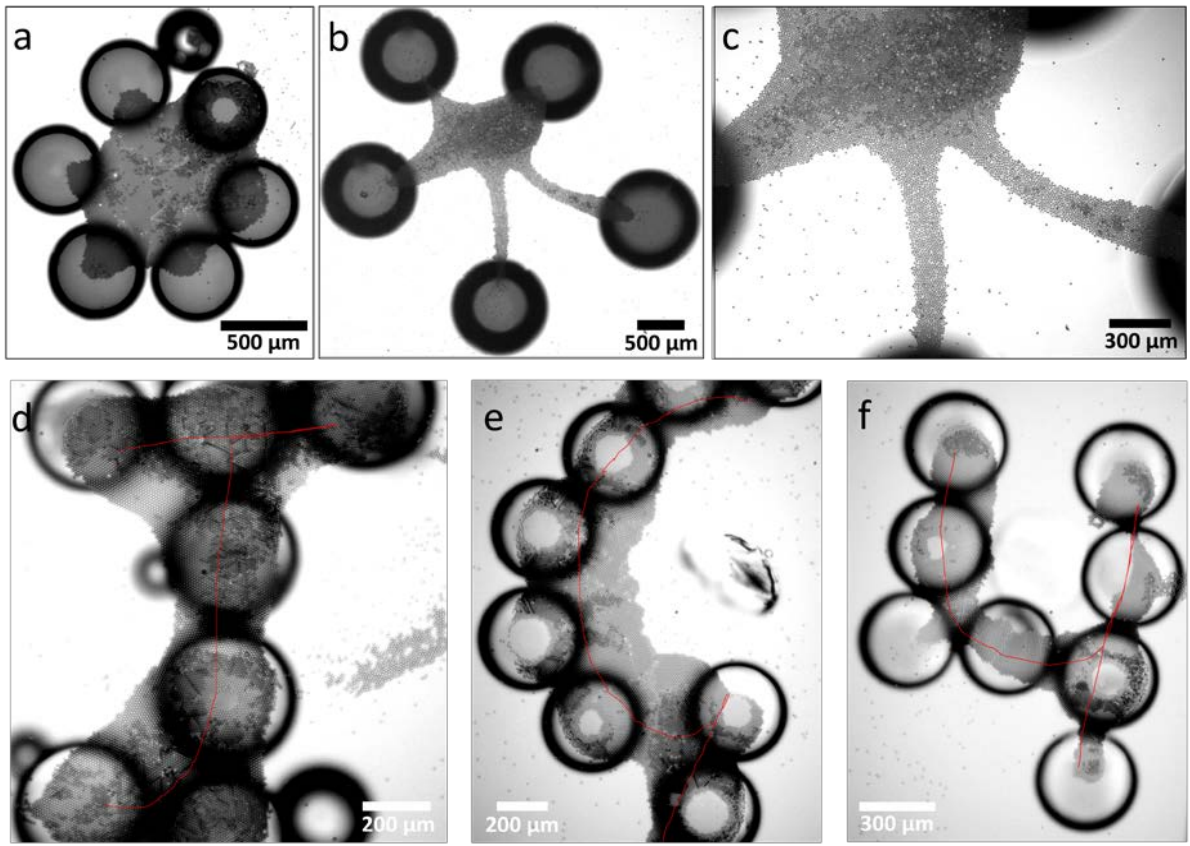

**Fig. 6** Preliminary results for patterning colloidal crystals utilizing IEX500-based inverted pumps: hexagon from PS10 particles at $H$= 0.5 mm (a), star from PSPVP5 particles at $H$= 1 mm (b and c), and logo of 'JGU' (Johannes Gutenberg University) from PS10 particles at $H$= 0.5 mm (d-f). Scale bars are as indicated.

Learning from manipulating the orientation of colloidal crystals from two IEX-based inverted pumps, we further patterned crystalline structures from colloidal particles utilizing several IEX-based inverted pumps. In Fig. 6a, the colloidal crystal forms the shape of hexagon with each angle pointing towards one projected center of the IEX. In Figs. 6b and 6c, colloidal crystal of star shape forms inside the region with weak flow due to the overlapped pH gradient. For this pattern, crystals nucleate beneath each IEX, followed by growing along the line connecting the center of the pentagon and the IEX. However, the center of the star has two or three layers as some of the small sized colloids are convected to the upper layer. To recover monolayer structure, large sized PS15 were used to assemble pentagon shaped crystal, which is however multi-domain (Fig. S5). In Figs. 6d-6f, we show a first attempt to grow colloidal crystals mimicking the shape of our university logo 'JGU' (Johannes Gutenberg University).

**Discussion**

We have assembled millimeter sized monolayer single colloidal crystals with desired orientations at pre-determined locations and even patterned structures with crystalline order without seed. Due to the long-ranged (in the radial range of millimeter) flow generated by the IEX, the colloidal assembly can be realized over a large area. As compared to previous efforts to assemble colloidal crystals using eo-pumps, the presently formed crystals are therefore much larger and show an exceptionally low defect density, the latter effect resulting from the assembly without seeds. Avoiding incommensurability with seed shape and size as well retaining particle mobility during the initial stages of assembly has proven to be the key to large-scale high quality crystals. Moreover, exploiting the stable symmetry breaking of the flow pattern when two IEX are present, we have been able to grow crystals at desired locations with pre-determined orientations. The simultaneous realization of location and orientation control without seed on an unstructured substrate appears to be unique for the present approach.

In principle, any kind of colloidal particles can be assembled with the right combinations of the ion-exchange process, effective charge of colloidal particles and the substrate. Thus, the inverted micro-fluidic pumps can be used to assemble different kinds of colloids such as $SiO_2$, $TiO_2$, Au and metal oxide particles. Different from the eo-pumping on a coated substrate, there is no sorting of colloids according to their size in the inverted pumps (video 5), due to the dominance of the convective flow on a highly charged substrate. However, this is not a drawback of the method, as mixing colloids of different kinds leads to more defects in the crystal, which is not desirable.

The principle restriction of applicability is given by the fact that we are working on the verge of the colloidal domain. The size range of colloidal particles which can be assembled is limited. For rather small sized colloids with diameter smaller than 500 nm, the disturbance of Brownian motion makes it hard to form crystals. Conversely, the flow friction is not able to drag rather big sized colloids ($2a > 60$ μm) towards the IEX. Changing height qualitatively balances with tuning of gravity and buoyancy matching. For example, particles of diameter 1-2 μm are assembled into multilayer crystalline structures at $H$= 0.5 mm due to the upward convective flow. To recover monolayer structure, one could increase the density of particles, e.g., by embedding $TiO_2$ nanoparticles or ferromagnetic nanoparticles, or decrease the fluid flow by increasing the cell height. Multilayer experiments are under way.

Concerning the mechanism of the fluid flow on the bottom substrate, the main transport mechanism for the colloids should be of hydrodynamic nature. In fact, it can readily be influenced via its geometry dependence. When the size of IEX is close to the cell height, strong confinement induces a qualitative change of the radial flow velocity decaying from $1/d$ to a $1/d^2$. It is, however, not clear whether transport in addition is influenced by a pH-gradient. Photo-optical experiments for pH detection can not discriminate layering effects and return only the height averaged pH. A qualitative check for a pH-gradient at the lower substrate was performed using melamine–formaldehyde (MF) resin particles, which change the sign of their surface zeta potential from positive to negative when the environmental pH is changed from acidic to alkaline.[38] At $H$= 1.5 mm, MF particles can be dragged to the projected center of the IEX500. While, at $H ≤ 0.8$ mm, the MF particles are drifted by the flow and stop approaching at some distance from the projected surface of the IEX500, indicating a local low pH induced positive charge of particles sticking to the negatively charged substrate. This demonstrates the presence of a radial pH gradient, which at present, however, has not been characterized any further. A detailed investigation on the mechanism of the flow is beyond the scope of this work but under way in our lab. A related question is the optimum ion concentration for crystal growth. We find that the low ionic condition employed here (deionized water with saturated atmospheric $CO_2$ inducing a background pH of ~5.5 and μmolar impurity cationic from the substrates[35,37]) is the optimum condition, under which colloidal transport is fast and colloids keep slight distance between them to allow rearrangement. Moreover, the screening from the low concentration background ions and self-screening of colloids plus the converging flow are able to stabilize the crystalline structure. Further increase the background salt concentration decreases the colloid-colloid distance and their ability to rearrange, leading to increased amount of defects. We further note that to tune the time scale for crystal growth, one can change the cell height, surface charge of the substrate, colloidal concentration, and size of the IEX. Systematic measurements at varied boundary conditions are under way.

Our observations naturally call for extensions. Speculation on future applications, first of all, it benefits many fundamental investigations. For example, the easy control of crystal orientation allows for the study of grain boundary motion and diffusion in and across grain boundaries with predetermined intersection angles. They may further allow studies of the roughening transition or studies of orientation dependent interfacial free energies.[39] For decoration purpose, Laue scattering may be exploited to obtain colored mosaics with color-play depending on viewing angle. Furthermore, it is in principle possible to extend the approach to other micro-fluidic pumps which allow geometric inversion, e.g., catalytic pumping.[40] It is also interesting to see how anisotropic particles, e.g., rod or cubic particles, are ordered in eo-pumps and the inverted pumps with growing interest in assembling anisotropic colloidal particles.[41-43] Moreover, it is feasible to implement some functionalities or switch to the inverted pumps at increased expenses. For instance, fixing $TiO_2$ particles on the top substrate, a reversible assembly and disassembly controlled by the on/off of UV light is accessible.[40] Additionally, the assembly range and speed can be manipulated in situ by changing the intensity of light. In the long run, combined with the advanced 3D printing technique,[44] micro crystalline structures of defined orientation may be obtained. Immobilizing the crystalline structure by photoinduced gelation[15,16] or heat induced inter-particle-crosslinking,[45] photonic materials for spectroscopy,[15,46] structured substrates for sensors[47,48] and colloidal epitaxy,[49] and matrices for optical composites[50] could be obtained.

**Conclusion**

We have demonstrated a simple seedless approach to fabricate millimeter sized monolayer single colloidal crystals at pre-determined locations using the ion exchange resin-based inverted microfluidic pumps. This approach allows easy manipulation of crystal orientation and therefore indicates possibilities for patterning structures on demand. The approach demonstrated is not restricted to the use of eo-pumps. In principle, any other pump type which allows geometric inversion can also be implemented to realize e.g. an on/off switch and other functionalities. Therefore, we expect our work to provide a useful approach for designing colloidal crystals for a wide range of applications.

**Experimental**

The main part of the study was carried out using cationic ion exchange resin particles of diameter 300-1200 µm (Amberlite K306, Roth GmbH, Germany), which were manually sorted into diameter classes of 400 ± 10 μm, 500 ± 10 μm and 660 ± 10 μm, denoted as IEX400, IEX500 and IEX660, respectively. Microgel-based cationic ion exchange resin of diameter 45 μm (lab code IEX45, CGC50×8, Purolite Ltd, UK) was used for the pH map. Most of the model colloidal particles were negatively charged polystyrene (PS) spheres (MicroParticles GmbH, Germany) of different diameters as determined by the manufacturer using electron microscope (lab codes PS5, PS10 and PS15). Home-made PVP stabilized PS particles of diameter 5 μm (lab code PSPVP5) were also used for colloidal assembly. Some melamine–formaldehyde (MF) resin particles of diameter 5 μm (denoted as MF5, Microparticles GmbH, Germany) were used to qualitatively test the local pH on the bottom substrate. Before use, all the colloidal suspensions were diluted (1:1000 v/v) with doubly distilled water from the stock suspension and deionized (Amberlite K306, Roth GmbH, Germany).

The sample cell was constructed from circular Perspex rings with diameter of $D$ = 20 mm attached to a microscopy slide by hydrolytically inert epoxy glue (UHU plus sofortfest, UHU GmbH, Germany) and dried for 24 h before use. Standard ring heights were $H$ = 0.5 mm, 1 mm and 1.5 mm. Commercial soda lime glass slides of hydrolytic class 3 (VWR International, Germany) served as top and bottom substrates. Before use, these were washed with 1% alkaline solution (Hellmanex®III, Hellma Analytics) under sonication for 30 min, then rinsed with tap water and subsequently washed with doubly distilled water for several times. Substrate surface potential under thoroughly deionized condition determined by Doppler velocimetry is $\zeta$ = -110 mV.[35]

The IEX particles were placed manually on the top glass slide at arbitrarily chosen locations and fixed using tiny amount of epoxy glue, then they were dried overnight before use. In the experiments, 200 μL of dilute colloidal suspension was added into the sample cell. Then the sample cell was quickly covered by the glass slide with the IEX particles fixed on it.

For pH measurements, a small amount of Universal indicator mixture (pH 0-5 and pH 4-10, 1:5 v/v, Sigma Aldrich, Inc.) was added into the sample cell of $H$= 0.5 mm. Images of RGB color were taken using a consumer DSLR (D700, Nikon, Japan) mounted on an inverted scientific microscope (DMIRBE, Leica, Germany). We first measured the calibration curve using buffer solutions of different fixed pH, which shows a monotonic decrease of blue-to-red color ratio with increasing pH.[37] Then applying the calibration curve to the images of samples, we obtained the pH values at each pixel around the IEX.

Samples were observed typically at 5× or 10× magnifications using an inverted scientific microscope (DMIRBE, Leica, Germany) connected to a standard video camera with a 0.5x lens to yield a large field of view. Videos of different frame rates (mainly 2 s/frame, 3 s/frame and 30 s/frame) were taken. For the pump flow velocity, tracer particles were tracked through extracting the perimeter using a home-written Python script. The velocity curves were averaged over 12 IEX particles and 160 tracer particles.

For crystal analysis, the videos were first processed by Image J to adjust the brightness and contrast, then analyzed using home-written python script. Particles were identified as in the crystalline phase using the $\phi_6$ bond order parameter defined as[51, 52]

$$\phi_6 = \frac{1}{N_b} \sum_{n=1}^{N_b} e^{6i\theta_{mn}}$$

Where $\theta_{mn}$ is the angle between the bond joining the inspected particle m and a nearest neighbor n and a fixed axis. Thresholds were set to have a number of next neighbor particles (defined as center-to-center distance less than 9.75 μm) $N_b \geq 2$ and $\phi_6 \geq 0.8$. Local crystalline orientations were obtained by fitting an ideal hexagonal nearest neighbor distribution to the actual nearest neighbor configuration. The orientation angle was taken as the minimum clockwise angular deviation between a nearest neighbor bond of the fitted pattern and the x-direction.

**Acknowledgment**

We are pleased to thank Joost de Graaf for the intense and helpful discussion about the pumping mechanism. We further thank Gabriele Schaefer for providing the PSPVP5 particles, Alexander Reinmüller, Hannah Müller for initiating the crystal printing research, Christopher Wittenberg for technical support in particle tracking programming and Julian Weber for the support in pH-measurements. Financial support of the DFG (SPP 1726, Grant No. Pa459/18-1,2) is gratefully acknowledged.

# Supplementary information

# Seedless assembly of colloidal crystals by inverted micro-fluidic pumping

Ran Niu*, Thomas Palberg

*Institut für Physik, Johannes-Gutenberg Universität, Staudingerweg 7, 55128, Mainz*

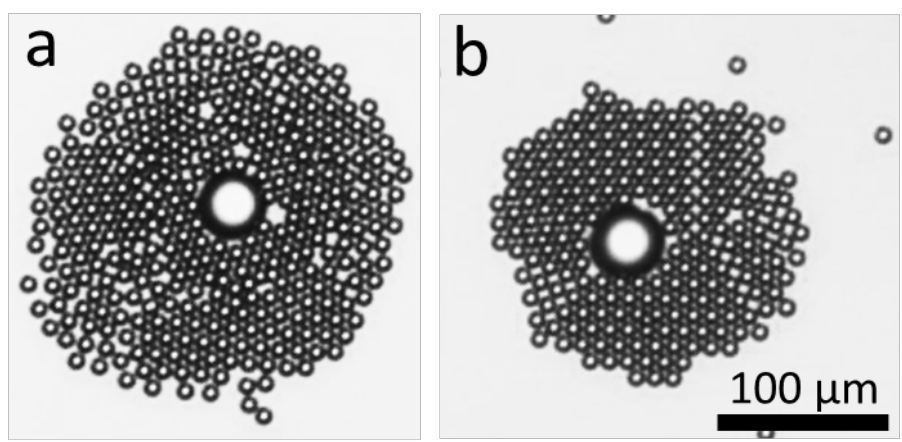

**Fig. S1** Low quality colloidal crystal of PS10 in eo-pumping of IEX45: no crystalline order (a) and multi-domain crystal on uncoated substrate due to the incommensurate of the seed and the formed crystal (b). The scale bar in (b) applies for both images.

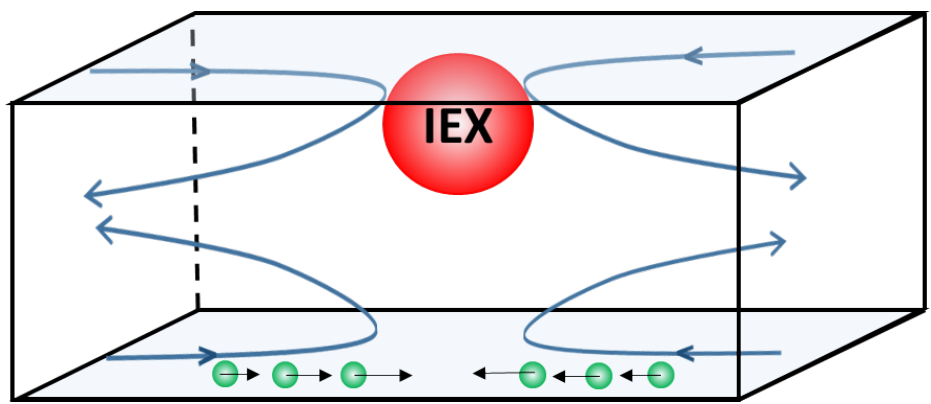

**Fig. S2** A schematic drawing of the IEX-based inverted microfluidic platform.

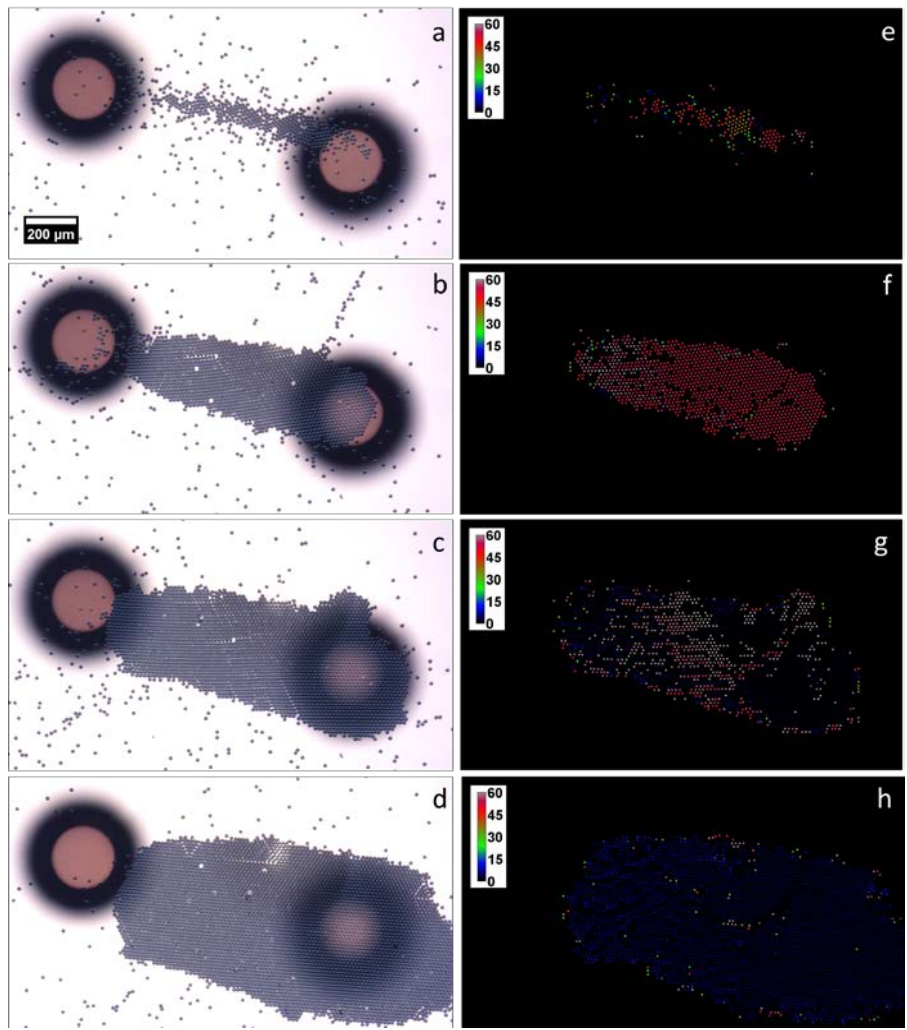

**Fig. S3** Typical stages of oriented crystal formation: PS15 in the inverted pump formed by two spaced IEX500 with center-to-center distance of 1116 µm and orientation of 16.9° relative to the horizontal axis. 200 µm scale bar shown in a) applies to all images. (a) At $t$ = 52.5 min, crystal forms between the two IEX500. (b-d) Crystal grows longer along the central line connecting the two IEX; meantime, the crystal grows wider in the direction perpendicular to the central line. In the process, dislocations and grain boundaries develop and anneal. Images were taken at $t$ =122.5 min, 167.5 min and 242.5 min, respectively. The corresponding maps of crystalline particles with $\phi_6$ > 0.8 and the color coded local crystalline orientation relative to the horizontal axis (e-h).

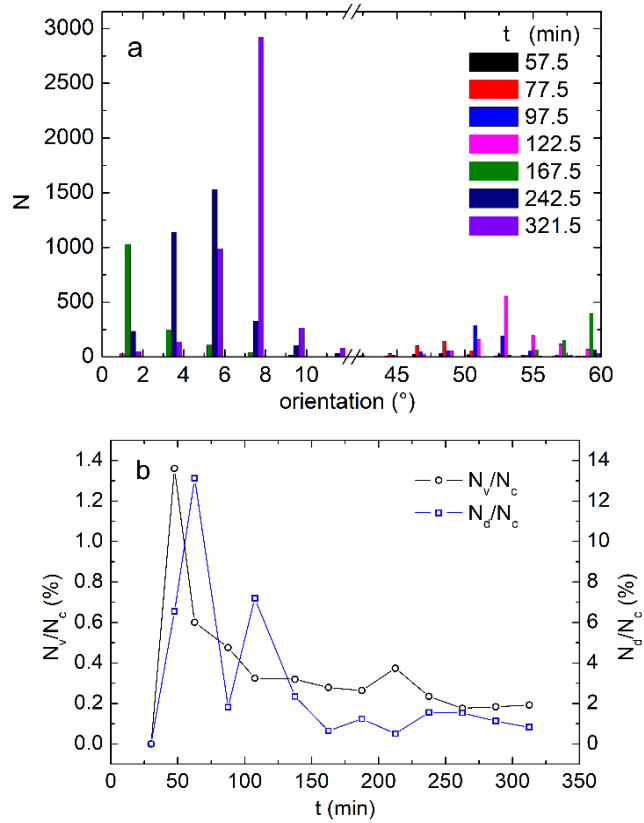

**Fig. S4** (a) Histogram of number of crystalline particles with certain orientations relative to the horizontal axis for the images shown in Fig. S3. (b) Fraction of voids $N_v/N_c$ and fraction of particles in dislocations and grain boundaries $N_d/N_c$ as a function of time $t$ from the start of the experiment.

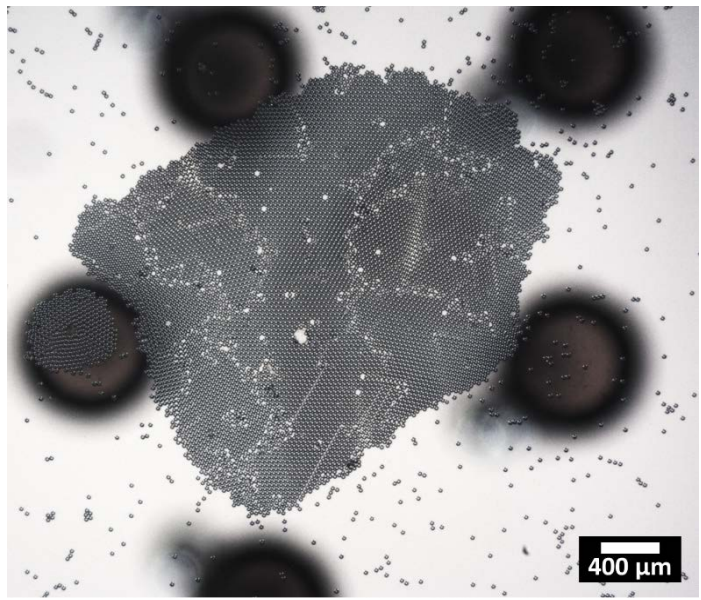

**Fig. S5** Patterning pentagon shaped monolayer colloidal crystal using IEX500-based inverted pump from PS15 at *H*= 1 mm. Scale bar as indicated.

**Videos:**

Video 1. Single IEX660-based inverted pump at *H*= 1 mm. PS7 tracer particles are dragged towards the projected center of IEX660. 14x real time speed. Scale bar as indicated.

Video 2. Single crystal formed in single IEX500-based inverted pump from PS10 at *H*= 1 mm. 600x real time speed. Image size 1217.0x971.6 µm$^2$.

Video 3. Oriented single crystal formed between two spaced IEX500 (center-to-center distance of 1487 µm, aligned along 30° relative to the horizontal axis)-based inverted pump from PS15 at *H*= 1 mm. 300x real time speed. Image size 1790.2x1342.6 µm$^2$.

Video 4. Oriented single crystal formed between two spaced IEX500 (center-to-center distance of 1116 µm, aligned along 16.9° relative to the horizontal axis)-based inverted pump from PS15 at *H*= 1 mm. 300x real time speed. Image size 1790.2x1342.6 µm$^2$.

Video 5. Assembly of PS15 and PS10 in the single IEX-based inverted pump. The assembled colloids are mixed without obvious sorting of the size. 200x real time speed. Image size 2727.3x2045.5 µm$^2$.